\makeatother

\documentclass[twocolumn, tighten]{aastex631}
\usepackage{graphicx}
\usepackage{listings}
\usepackage{rotating}
\usepackage{ulem}
\usepackage{stix2}

\newcommand{\lapprox} {\, \lower3pt\hbox{$\sim$}\llap{\raise2pt\hbox{$<$}}\,}
\newcommand{\gapprox} {\, \lower3pt\hbox{$\sim$}\llap{\raise2pt\hbox{$>$}}\,}

\begin{document}


\title{The efficiency of electron acceleration during the impulsive phase of a solar flare}

\author[0000-0002-8078-0902]{Eduard P. Kontar}
\affiliation{School of Physics \& Astronomy, University of Glasgow, Scotland, UK}

\author[0000-0001-8720-0723]{A. Gordon Emslie}
\affiliation{Department of Physics \& Astronomy, Western Kentucky University, Bowling Green, KY 42101, USA}

\author[0000-0001-7856-084X]{Galina G. Motorina}
	\affil{Astronomical Institute of the Czech Academy of Sciences, 251 65 Ond\v{r}ejov, Czech Republic\\
Central  Astronomical Observatory at Pulkovo of Russian Academy of Sciences, St. Petersburg, 196140, Russia\\
Ioffe Institute, Polytekhnicheskaya, 26, St. Petersburg, 194021, Russia}

\author[0000-0001-8585-2349]{Brian R. Dennis}
\affiliation{Solar Physics Laboratory, Code 671, Heliophysics Science Division, NASA Goddard Space Flight Center, Greenbelt, MD 20771, USA}

\begin{abstract}
Solar flares are known to be prolific electron accelerators, yet identifying the mechanism(s) for such efficient electron acceleration in solar flare (and similar astrophysical settings) presents a major challenge. This is due in part to a lack of observational constraints related to conditions in the primary acceleration region itself. Accelerated electrons with energies above $\sim$20~keV are revealed by hard X-ray (HXR) bremsstrahlung emission, while accelerated electrons with even higher energies manifest themselves through radio gyrosynchrotron emission. Here we show, for a well-observed flare on 2017~September~10, that a combination of \emph{RHESSI} hard X-ray and and SDO/AIA EUV observations provides a robust estimate of the fraction of the ambient electron population that is accelerated at a given time, with an upper limit of $\lapprox 10^{-2}$ on the number density of nonthermal ($\ge 20$~keV) electrons, expressed as a fraction of the number density of ambient protons in the same volume. This upper limit is about two orders of magnitude lower than previously inferred from microwave observations of the same event. Our results strongly indicate that the fraction of accelerated electrons in the coronal region at any given time is relatively small, but also that the overall duration of the HXR emission requires a steady resupply of electrons to the acceleration site. Simultaneous measurements of the instantaneous accelerated electron number density and the associated specific electron acceleration rate provide key constraints for a quantitative study of the mechanisms leading to electron acceleration in magnetic reconnection events.
\end{abstract}

\section{Introduction}\label{introduction}

Spatially resolved hard X-ray (HXR) observations of solar flares reveal that electron acceleration likely occurs in magnetic reconnection outflow regions \citep[see, e.g.,][for reviews]{2011SSRv..159..107H,2017LRSP...14....2B}. Indeed, observations from the Ramaty High Energy Solar Spectroscopic Imager \citep[\emph{RHESSI};][]{2002SoPh..210....3L} observations have demonstrated \citep{2003ApJ...596L.251S,2007AdSpR..39.1398A,2008ApJ...676..704L,2013ApJ...779..107B,2014ApJ...780..107K,2019ApJ...872..204B} the presence of faint sources of HXR emission above the tops of coronal loop structures as evidence for magnetic reconnection, as first noted by \citet{1994Natur.371..495M}. These sources are consistent with regions where magnetic turbulence could provide sufficient power to accelerate electrons to the required energies \citep{2017PhRvL.118o5101K,2019ApJ...877L..11R,2020ApJ...890L...2C,2021ApJ...923...40S}. It is well known \citep[e.g.,][]{1997JGR...10214631M,2011SSRv..159..357Z} that the total number of electrons accelerated during a large solar flare can exceed the total number of electrons in the coronal source region, so that continuous resupply of the acceleration region \citep[e.g., by a cospatial return current;][]{1977ApJ...218..306K,1980ApJ...235.1055E,1997A&A...320L..13Z, Alaoui_2017} is an essential ingredient of a viable electron acceleration model.

 \textit{Spatially-integrated} bremsstrahlung hard X-ray emission, produced during collisions of non-thermal electrons on ambient ions, provides direct diagnostics of the \textit{total} electron acceleration rate $\dot{N}$ (s$^{-1}$) at energies above a specified energy (typically taken to be 20~keV). In such a ``thick-target'' calculation, the deduced electron acceleration rate is generally independent of the density structure of the flaring region, but the relationship between the accelerated electron distribution and the emitted HXR spectrum does depend on whether a cold \citep{1971SoPh...18..489B,1972SvA....16..273S} or warm \citep{2015ApJ...809...35K} thick target model is used. (The dependency of total acceleration rate arises through the difference in the energy loss term  and transport through plasma \citep{2019ApJ...871..225K} -- where the average energy of the ambient electrons is much smaller than the energy of the accelerated electrons  -- versus warm targets, where the energies of the accelerated and target particles can be comparable.) The inferred total electron acceleration rate depends strongly on the value of the low-energy spectral cutoff, which is difficult to determine from observations \cite[e.g.][as a review]{2019ApJ...881....1A}.

By contrast with the spatially-integrated case, the HXR flux spectrum from a given spatial sub-region in the flare provides information on the density-weighted mean electron flux spectrum in that region. For this analysis, ``thin-target'' modeling \citep{2003ApJ...595L.115B} is used; this does not require any assumptions about the nature of the prevailing electron transport mechanism. This density weighting implies that bright hard X-ray footpoints are conspicuously evident \citep[e.g.,][]{2003ApJ...595L.107E,2008A&A...489L..57K} because of the high density in the chromosphere, while strong coronal HXR sources are relatively rare \citep{2011SSRv..159..107H} because the density is so low.  Given the limited dynamic range of \emph{RHESSI} and similar instruments that employ indirect (Fourier-transform based) imaging techniques, the problem of characterizing the HXR emission from low-density coronal acceleration regions is exacerbated by the presence of much brighter footpoint emission in the field of view.

Of considerable interest in understanding the nature of the process that accelerates particles to high energies are the ratios of the number densities (cm$^{-3}$) of nonthermal and thermal electrons ($n_{nth}$ and $n_{th}$, respectively) to the total number density of background electrons, which, from considerations of charge neutrality, is equal to the number density of protons in the same region: $n_p = n_{th} + n_{nth}$ in hydrogen plasma. The instantaneous fraction of accelerated electrons in the flaring corona has been previously discussed in the literature, but the results, methods, and the region that defines the ``coronal source'' and ``above-the-looptop'' differ somewhat and hence necessitates addition investigations.  \cite{2010ApJ...714.1108K} reported $n_{nth}(>\text{16~keV})/n_p \simeq 0.1$, while \citet{2013ApJ...764....6O} re-analysed the flare and deduced $n_{nth}(>$$20$~keV)$/n_p \simeq 0.01$ (see their Equation~(4)). \cite{2014ApJ...780..107K} analysed a limb flare in the presence of bright footpoint emission and found $n_{nth}(>(10-15)~\text{keV})/n_p \simeq 1$ in the region above the coronal source, assuming a power-law spectrum with a sharp low-energy cut-off in the range from $10-15$~keV.  However, this high value was obtained using a Maxwellian+power-law fit to the hard X-ray spectrum. Mean source electron spectra \citep{2003ApJ...595L.115B} in some coronal hard X-ray sources have been found to be more consistent with a kappa distribution \citep{2009A&A...497L..13K}, which rolls over smoothly from a power-law at high energies to a Maxwellian at low energies. Compared to fitting the thermal and nonthermal (power-law) parts of the spectrum separately, fitting the data with a kappa distribution generally results \citep[as in][]{2013ApJ...764....6O} in a significant reduction in the required electron flux near the rollover energy and hence a lower inferred value of $n_{nth}$. Indeed, \cite{2015ApJ...815...73B} demonstrate that using a kappa-distribution fit to \emph{RHESSI} data, while also incorporating EUV emission line data to constrain the fit, results in a reduction in $n_{nth}$ by a factor of up to $\sim$30 compared with the commonly-used power-law-only fit. Early in a flare, electron distributions in magnetic reconnection outflow regions could form a bulk thermal distribution with a relatively steep (e.g., power-law) non-thermal tail, so that the electrons above, say, $20$~keV represent only a small fraction of the total ambient population \citep{2019ApJ...872..204B}. However, the spectrum could change during the peak of the flare. Radio emission produced by gyrosynchrotron radiation of accelerated electrons with relatively high energies $\gapprox 100$~keV is also commonly observed in the flaring atmosphere \citep[see][for reviews]{1982ApJ...259..350D,2020FrASS...7...57N}
and provides insight into the properties of electrons with energies higher than those that generate most of the HXR emission \cite[e.g.][]{2011SSRv..159..225W,2018A&A...610A...6M}.

In this paper, we combine thin-target and thick-target modeling of \emph{RHESSI} X-ray observations of a well-observed solar flare with contemporaneous EUV observations from the Solar Dynamics Observatory Atmospheric Imaging Assembly \cite[SDO/AIA;][]{2012SoPh..275...17L} in order to better constrain both the total number of accelerated electrons and the all-important ratio $n_{nth}/n_p$. The flare occurred on 2017~September~10 and revealed clear evidence \citep{2020NatAs...4.1140C} for a reconnection current sheet located above the flare loop-top. \cite{2022Natur.606..674F} have shown from Expanded Owens Valley Solar Array \citep[EOVSA;][]{2018ApJ...863...83G} radio observations of this event that there was a relative dearth of thermal emission from an extended coronal volume but with considerable non-thermal emission from the same volume. They suggested that an inductive electric field of order 20~V~cm$^{-1}$ was generated by a magnetic field decay rate of order 5 G~s$^{-1}$ in a reconnection region of extent $\ell \simeq 4 \times 10^8$~cm, and concluded that almost \textit{all} of the electrons within that extended volume were accelerated to energies in excess of 20~keV, corresponding to $n_{nth}/n_p \simeq 1$ (and $n_{th}/n_p \simeq 0$). Our HXR and EUV observations conclusively demonstrate, however, that the instantaneous accelerated fraction $n_{nth}/n_p$ is much less than unity, specifically $\lapprox 0.01$. Moreover, we argue that the accelerating electric fields are likely considerably smaller in magnitude than suggested by \cite{2020Sci...367..278F} and \cite{2022Natur.606..674F}, indicative of a much smaller scale length $\ell \simeq 10^3$~cm associated with the reconnecting magnetic fields in the acceleration region.

\section{X-Ray and EUV Observations}\label{HXR_obs}

\begin{figure}[pht]
    \centering
    \vskip -5mm
    \includegraphics[width=.49\textwidth]{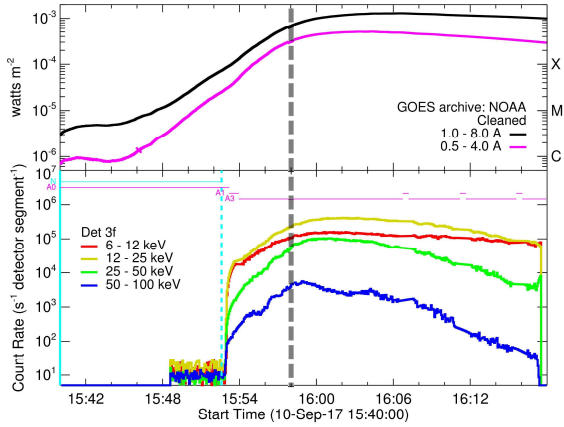}
    \vskip -7mm
    \caption{GOES and \emph{RHESSI} X-ray lightcurves of the X8.2 flare on 2017~September~10, in the indicated energy ranges. Note that \emph{RHESSI} data are only available from 15:52:30~UT at the start of the daytime part of the orbit until 16:17:30~UT on entry into the South Atlantic Anomaly. The horizontal purple lines indicate the times in the different attenuator states - A0, A1, and A3. The plotted count rates are shown for the front segment of Detector \#3 (3f), corrected to first order to the rates that would have been recorded in the A1 attenuator state, i.e., with only the thin attenuators in place. The dashed vertical line at 15:58~UT indicates the time interval of interest. Note that the periodic ripples in the lightcurves after about 16:06~UT are the result of uncorrected changes in the orientation of the instrument axis with respect to the spacecraft spin axis, caused by the attenuator motions \citep[see][]{2011A&A...530A..47I}.}
    \label{fig:hxr_lc}
\end{figure}
\begin{figure}[pht]
    \centering
    \vskip -5mm
    \includegraphics[width=0.49\textwidth]{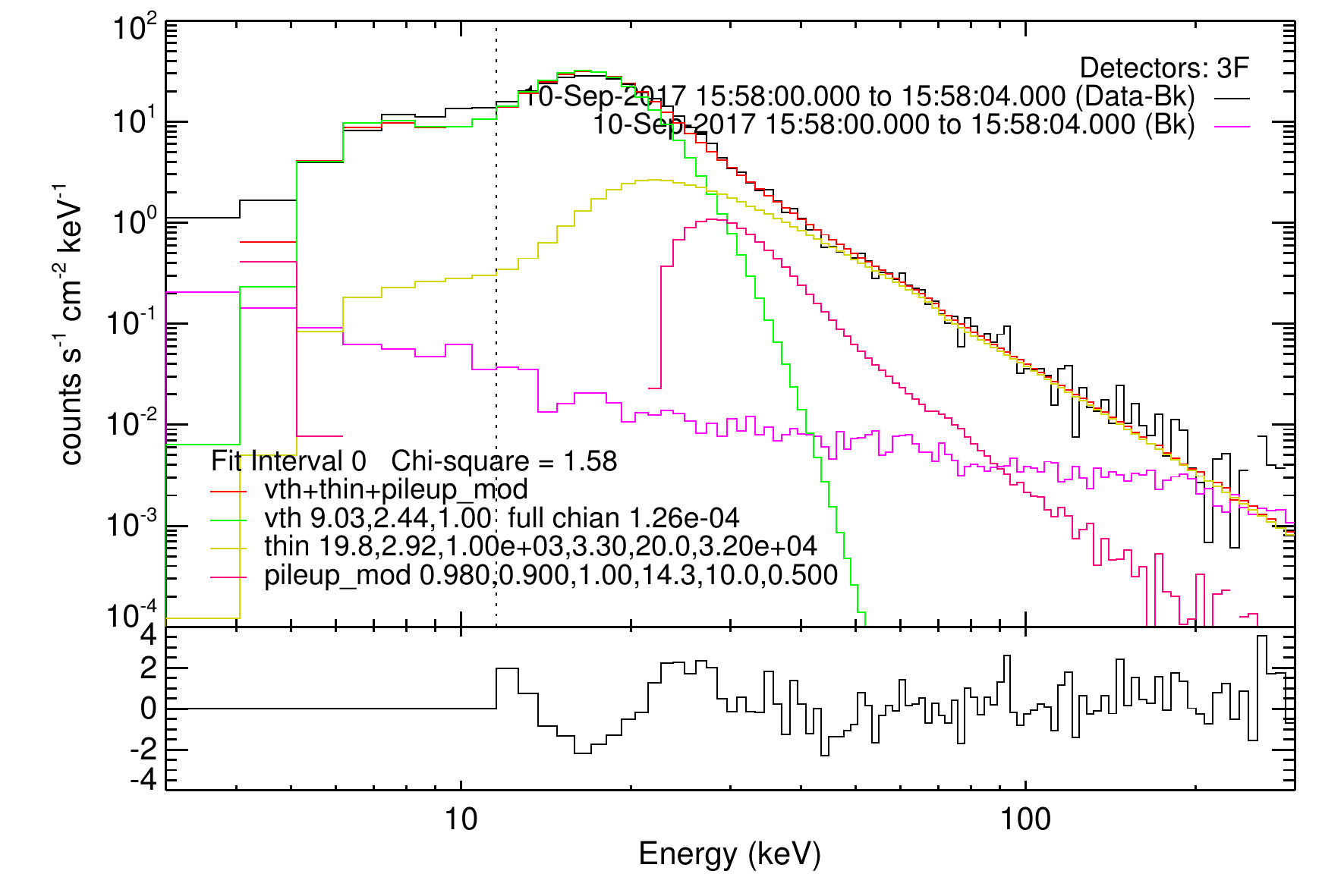}
    \vskip -5mm
    \caption{Spatially integrated hard X-ray (HXR) spectrum measured by Detector \#3, one of the four working \emph{RHESSI} detectors near the peak of the flare, in a 4~s interval starting at 15:58~UT. The HXR spectrum is well fitted by a combination of a thermal (``vth,'' green) and nonthermal (``thin,'' yellow) components. The ``thin'' function is used here for the nonthermal component assuming thin-target interactions of electrons with a power-law spectrum. The listed fit parameters are as follows: A(0) (here $19.8$) is the value of $\langle n_p V {\cal F}_o \rangle$ (in units of $10^{55}$~electrons~cm$^{-2}$~s$^{-1}$) at energies above reference energy $E_o$, A(1) is the electron power-law index $\delta$ (here 2.92), and A(4) is the value of $E_o$ (here 20~keV). The parameters A(2), A(3) and A(5) define the extended electron spectrum at energies above the range of interest in the present analysis, i.e., the upper limit (``break energy''; here $1000$~keV) to this power law, the spectral index above this break energy (here $3.3$), and the maximum energy for which the spectrum is computed (here $32$~MeV). The thermal and nonthermal contributions are comparable at an X-ray energy $\epsilon \simeq 27$~keV. The total spectrum (red) also includes the pileup component (lower red curve) that peaks near 30~keV; the background spectrum is shown in magenta. The bottom panel shows residuals from the fit in units of the the $1\sigma$~statistical uncertainties on the measured count rates at each energy.}
    \label{fig:hxr_fit}
\end{figure}

We consider the well-observed solar flare GOES X-class flare SOL2017-09-10T15:58, the GOES lightcurve of which is shown in Figure~\ref{fig:hxr_lc}. This flare appeared at the western limb of the solar disk, so that its vertical geometry can be well ascertained. The hard X-ray emission evolved over tens of minutes, but we here focus on a narrow time interval near the peak of the impulsive phase, around 15:58~UT. The spatially-integrated hard X-ray spectrum at this time (Figure~\ref{fig:hxr_fit}) shows both thermal and non-thermal components, with the spectrum dominated by non-thermal emission above about $30$~keV.

The spatially-integrated hard X-ray spectrum can be used to infer the total number of accelerated electrons above a chosen reference energy, and dividing this by the volume of the hard X-ray source (obtained from imaging spectroscopy data), this leads to a measure of the accelerated electron number density $n_{nth}$ (cm$^{-3}$). To quantify these relationships precisely, consider a region of volume $V$ (cm$^3$), populated by accelerated electrons with number density spectrum $n(E)$ (electrons~cm$^{-3}$~keV$^{-1}$). The associated electron flux $F(E)$ (electrons~cm$^{-2}$~s$^{-1}$~keV$^{-1}$) is $F(E) = n(E) \, v(E)$, where the (non-relativistic) electron speed $v(E) = \sqrt{2E/m_e}$, $m_e$ being the electron mass. This population of electrons produces a bremsstrahlung hard X-ray spectrum $I(\epsilon)$ (photons~cm$^{-2}$~s$^{-1}$~keV$^{-1}$) at a distance $R$ from the emitting region given by \citep{2003ApJ...595L.115B}

\begin{equation}\label{eq:Ieps}
  I(\epsilon)=\frac{1}{4\pi R^2} \, \int_{\epsilon}^{\infty} \langle n_p V F(E) \rangle \, \sigma(\epsilon, E) \, dE \,\,\, .
\end{equation}
Here $\sigma(\epsilon, E)$ is the bremsstrahlung cross-section\footnote{see OSPEX software \url{https://hesperia.gsfc.nasa.gov/rhessi3/software/spectroscopy/spectral-analysis-software/index.html}} . The quantity $\langle n_p V F(E) \rangle$ is the \textit{mean source electron spectrum}, the volume integral of the local electron flux (differential in energy), weighted by the volume-averaged number density of target protons $n_p$, and $\langle \cdots \rangle$ emphasizes that the quantity is a volume average. It is important to note that $\langle n_p V F(E) \rangle$ is directly determined from the observed hard X-ray spectrum, requiring only knowledge of the bremsstrahlung cross-section involved; no knowledge or assumptions regarding the source geometry or the physics governing the propagation of the accelerated electrons is required.

Because the mean source electron spectrum corresponds to the instantaneous distribution of high-energy electrons in the coronal target, its form is obtained by a ``thin-target'' fit to the bremsstrahlung hard X-ray spectrum.  Figure~\ref{fig:hxr_fit} shows that the instantaneous spectrum of nonthermal electrons that best fits the hard X-ray spectrum\footnote{thin-target fit using OSPEX \url{https://sohoftp.nascom.nasa.gov/solarsoft/packages/spex/doc/ospex_explanation.htm}} for this event is well-represented by a power-law of the form

\begin{equation}\label{eq:F}
F(E) = (\delta - 1) \, \frac{{\cal F}_o}{E_o} \, \left(\frac{E}{E_o}\right)^{-\delta} \,\,\, ,
\end{equation}
where $E_o$ is a reference energy (e.g., a low-energy cutoff), ${\cal F}_o$ is the total electron flux (electrons~cm$^{-2}$~s$^{-1}$) above that energy, and $\delta$  is the power-law spectral index.

From this follows the accelerated number density spectrum (electrons~cm$^{-3}$~keV$^{-1}$)

\begin{equation}\label{eq:nE}
n(E) = \frac{F(E)}{v(E)} = \sqrt{\frac{m_e}{2}} \,\, (\delta - 1) \, {\cal F}_o \, E_o^{\delta-1} \, E^{-\delta - 1/2} \,\,\, ,
\end{equation}
so that the number density of nonthermal electrons $n_{nth}$ (defined as the number of electrons per cm$^3$ that are accelerated to energies above $E_o$) is

\begin{equation}\label{eq:n-accelerated}
n_{nth} = \int_{E_o}^\infty n(E) \, dE = \frac{(\delta - 1)}{\left (\delta - \frac{1}{2} \right )}\, \frac{{\cal F}_o}{v_o} \, ,
\end{equation}
where $v_0=\sqrt{2E_0/m_e} \simeq 8 \times 10^{9}$~cm~s$^{-1}$~for~$E_0 = 20$~keV. The total mean source electron flux (electrons~cm$^{-2}$~s$^{-1}$ above energy $E_o$) may be written as the integral of the electron flux spectrum (Equation~(\ref{eq:F})) multiplied by the proton density and emitting volume, so that

\begin{equation}\label{eq:nVF}
\langle n_p V {\cal F}_o \rangle  = n_p \, V \, \int_{E_0}^{\infty} F (E) \, dE  \,\,\, .
\end{equation}
Combining Equations~(\ref{eq:n-accelerated}) and~(\ref{eq:nVF}), we find that

\begin{equation}\label{eq:nnth-np-V}
n_{nth} \, n_p = \frac{(\delta - 1)}{\left ( \delta - \frac{1}{2} \right )}\, \frac{\langle n_p V {\cal F}_o \rangle \, }{v_o \, V }  = \frac{(\delta - 1)}{\left ( \delta - \frac{1}{2} \right )} \, \sqrt{\frac{m_e}{2E_o}} \,\, \frac{\langle n_p V {\cal F}_o \rangle }{V } \,\,\, .
\end{equation}
The observationally-inferred value of $\langle n_pV {\cal F}_o \rangle $, coupled with estimates of the source volume $V$, thus straightforwardly constrains the product of the number of accelerated electrons $n_{nth}$ and the mean number density of target protons $n_p$ in the region under study.

\begin{figure*}[pht]
    \centering
    \includegraphics[width=0.49\linewidth]{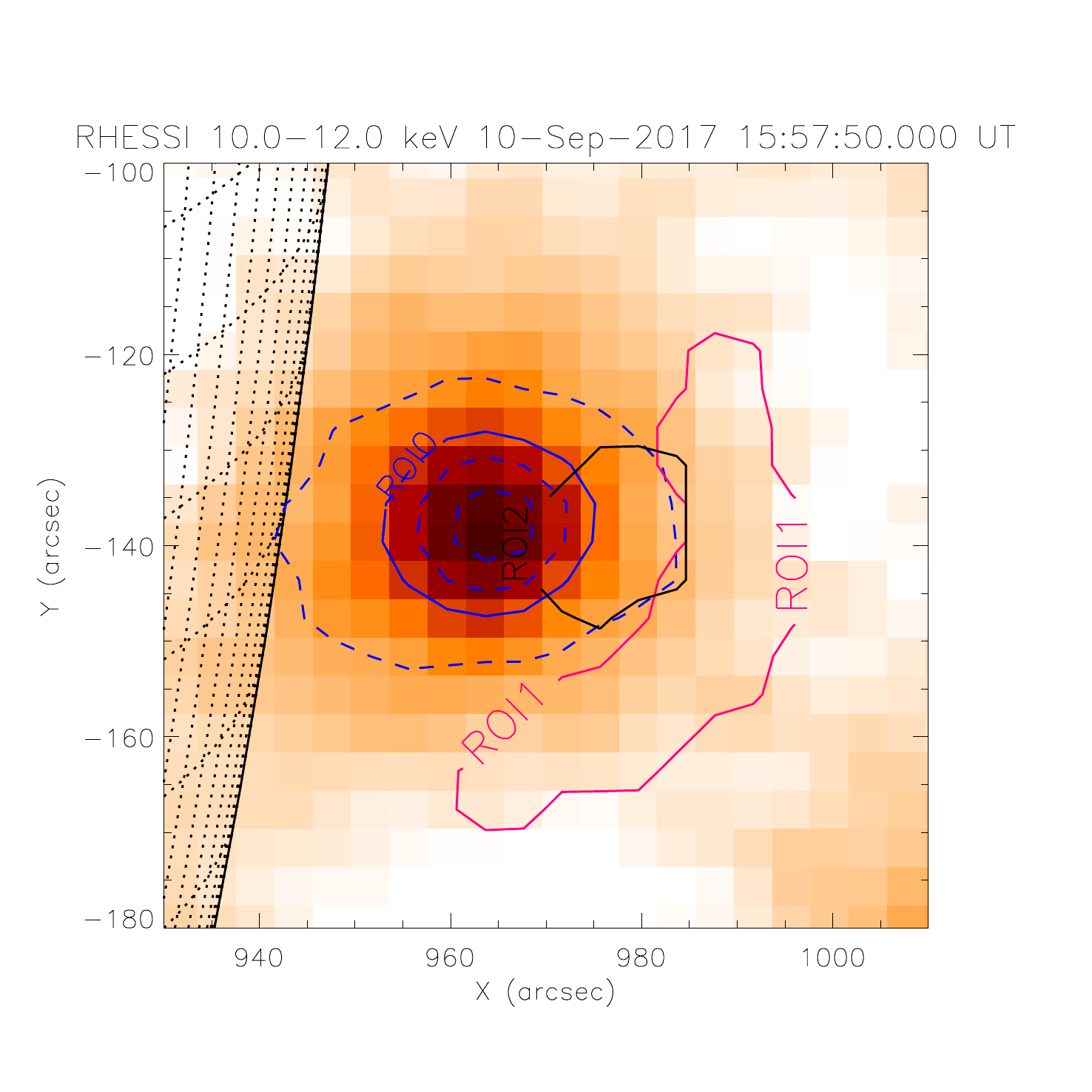}
    \includegraphics[width=0.49\linewidth]{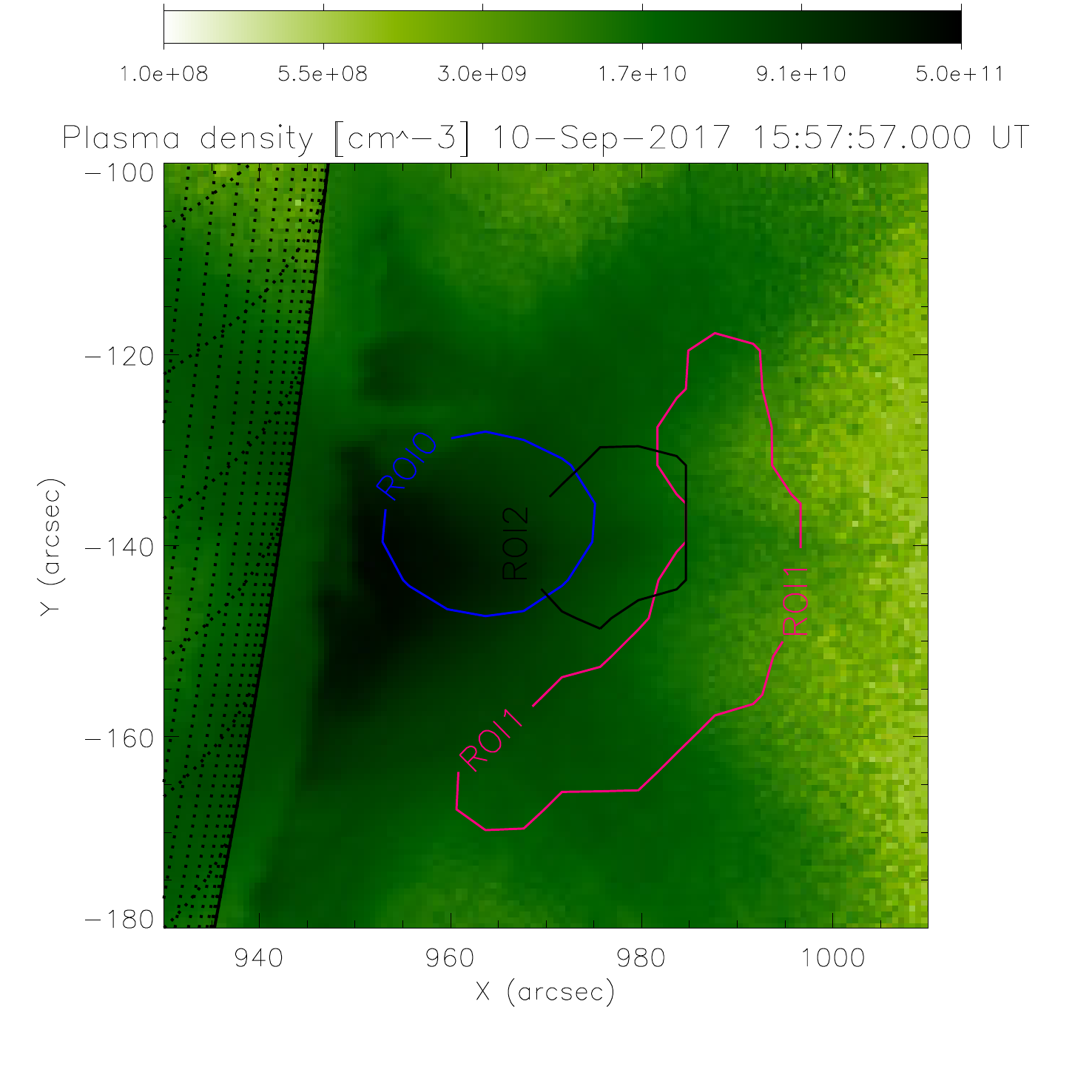}
    \vskip -5mm
    \caption{\textit{Left:} Clean \emph{RHESSI} X-ray images using Detectors \#3 and \#6 for the 20~s interval centered on 15:58~UT, with three regions of interest identified. The color-scale image represents thermal (10-12~keV) emission. The blue dashed lines are the contours at 20, 70, and 90\% of the peak 50-100~keV non-thermal emission. The region of interest labeled ROI-0 is defined by the solid blue contour at the 50\% level of the 50-100~keV image. Regions of interest ROI-1 and ROI-2 from \cite{2022Natur.606..674F} are depicted by solid black and magenta boundary lines, respectively. \textit{Right:} Thermal plasma density map, also showing ROI-1 and ROI-2. This was constructed by applying a regularized differential emission measure algorithm \citep{2012A&A...539A.146H} to SDO/AIA data for the event in question, assuming the same line-of-sight distance of $8^{\prime \prime}$ used by \cite{2022Natur.606..674F}.}
    \label{fig:hxr-50}
\end{figure*}

\begin{figure*}[pht]
    \centering
    \includegraphics[width=0.42\linewidth]{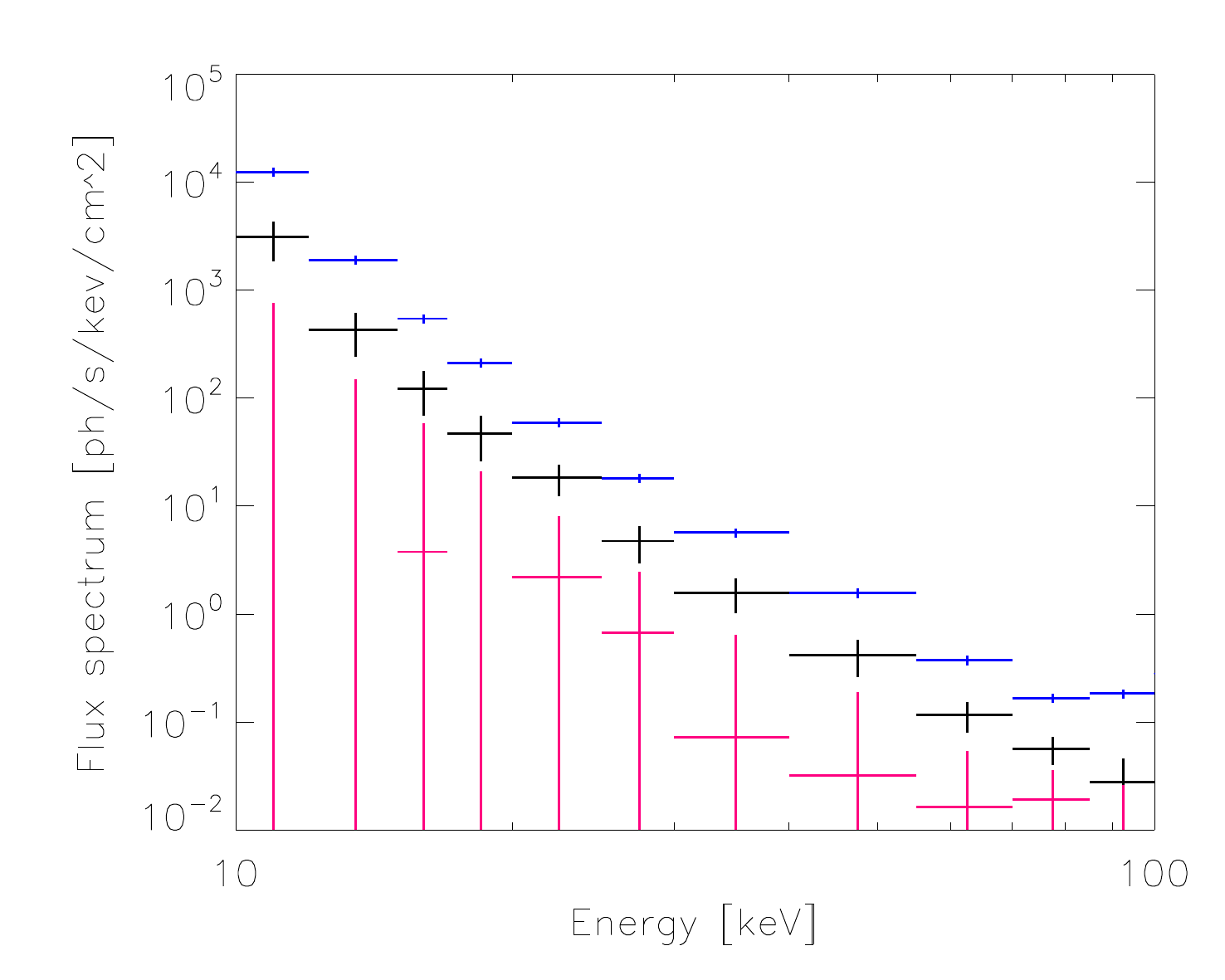}
    \includegraphics[width=0.42\linewidth]{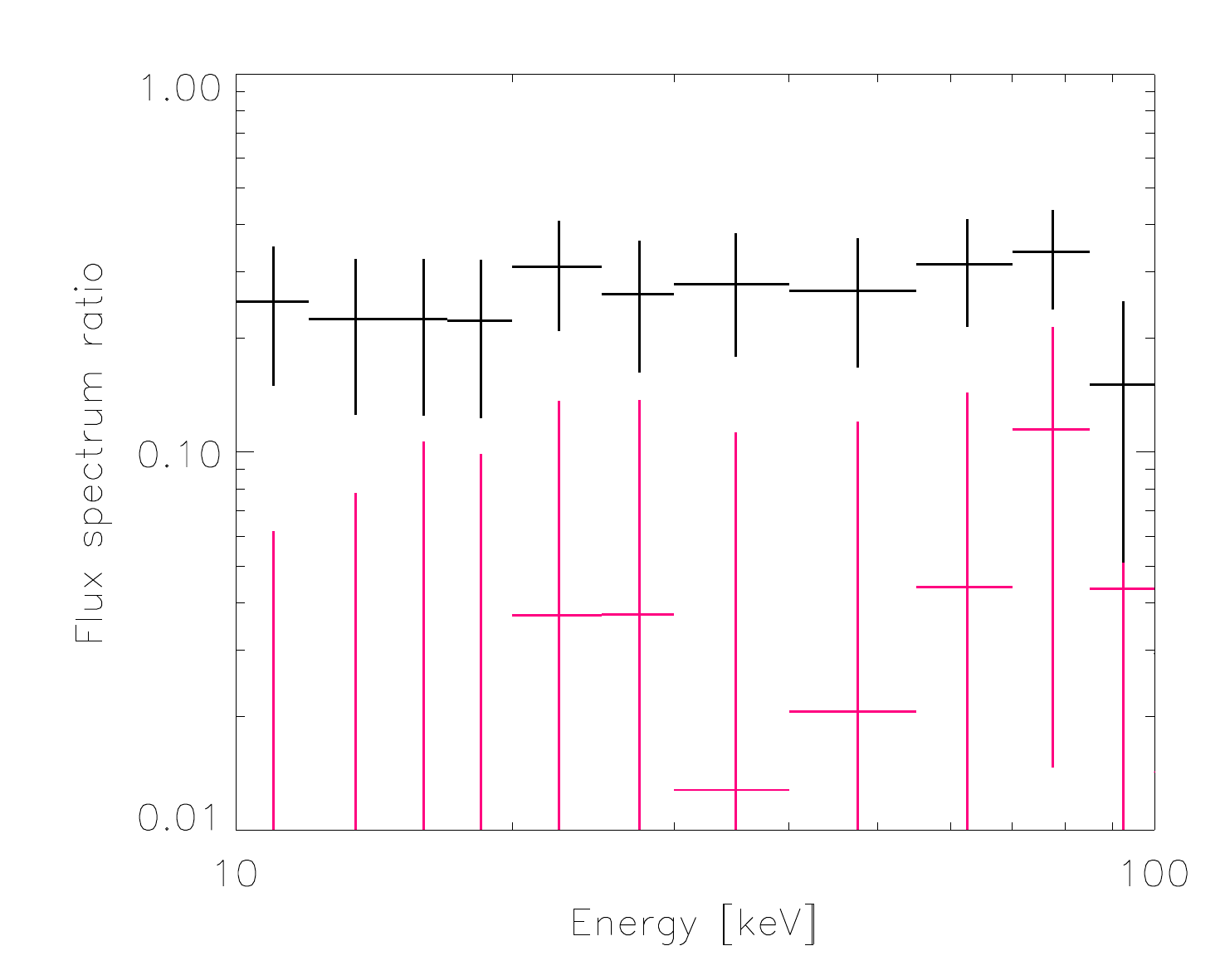}
    \vskip -3mm
    \caption{\textit{Left:} Hard X-ray flux spectrum in the selected regions: the integrated emission is shown in blue, the emission from ROI-1 is shown in magenta, and the emission from ROI-2 is shown in black; \textit{Right:} Electron flux spectrum in ROI-1 (magenta) and ROI-2 (black), normalized by the image-integrated spectrum.
    }
    \label{fig:xspectra}
\end{figure*}

\cite{2022Natur.606..674F} have identified two ``regions of interest'' shown in Figure~\ref{fig:hxr-50}, ROI-1 (magenta line) and ROI-2 (black line). ROI-1 is the extended curved elliptical region that outlines the volume in which they claim nearly 100\% electron acceleration efficiency. This claim is based on EOVSA imaging spectroscopy analysis showing that ROI-1 has (1) a large nonthermal electron population, (2) a relative dearth of thermal electrons over an extended (4~minute) period of time, and (3) a rapid magnetic field decay \citep[reported earlier by ][]{2020Sci...367..278F} that leads to a large induced electric field, capable of accelerating the entire ambient population in a very short time. For comparison, ROI-2 is identified as a region ``of more typical flare plasma, outside the acceleration region.''

X-ray imaging spectroscopy of this event can be made by analyzing \emph{RHESSI} observations of the same flare. Various image reconstruction algorithms, described by \cite{pianabook}, can be used such as Clean, MEM\_NJIT, uv\_smooth, and Vis\_FwdFit. The X-ray image in Figure~\ref{fig:hxr-50} was produced with the Clean method using observations over a 20~s interval centered at 15:58:00~UT. It reveals a source which we term ROI-0, with a FWHM area of $256$~arcsec$^2$ and a centroid at a slightly lower altitude (some 15 arcsec) than ROI-1 and ROI-2. This image shows that the vast majority of the hard X-ray emission at this time comes not from ROI-1, but rather from the compact ROI-0 source, the structure of which is rather poorly resolved by \emph{RHESSI}.

X-ray spectra for each region of interest, produced using the \emph{RHESSI} imaging spectroscopy technique described by \cite{2003ApJ...595L.107E}, \cite{2006A&A...456..751B}, \cite{2007ApJ...665..846P}, \cite{2008SoPh..250...53S}, and \cite{2014ApJ...787...86J}, are presented in Figure~\ref{fig:xspectra}. These imaging spectroscopy results are somewhat limited for two reasons: (1) only four of the nine \emph{RHESSI} germanium detectors (\#s 1, 3, 6, and 8, with FWHM angular resolutions of 2.3, 6.8, 35.3, and 106 arcsec, respectively) were operational at this time, and (2) the count rates were sufficiently high to cause significant pulse pileup, where two or more low-energy photons arriving within a short time ($\sim$$1~\mu$s) of each other are recorded as a single high-energy count. (One can see in Figure~\ref{fig:hxr_fit} a clear signature of pulse pileup at or below the $10-20\%$ level in the spectrum recorded by Detector \#3 near 30 keV; the effects of pile-up are weaker for other energies.) In general, at energies above the peak in the count-rate spectrum (here in the A3 attenuator state at $\sim$$18$~keV) pulse pileup tends to increase visibility amplitudes. This is especially true for the detectors with the coarser grids (\#s~6 and~8), since the signals from these detectors have the greatest modulation amplitudes and the pileup effect is greater at the peaks of the modulation cycles than in the valleys. Detector \#6 has the highest sensitivity of all the operating detectors, so that the estimated source extent at the characteristic 35-arcsec scale sampled by this detector is subject to an artificial increase, so the source could appear larger than the actual source extent and hence the X-ray intensity in both ROI-1 and ROI-2 could be overestimated.

Performing spatial integrations over the three regions of interest (ROI-0 through ROI-2) results in the three HXR source spectra shown in Figure~\ref{fig:xspectra}.  Similar to the methodology of previous studies \cite[e.g.,][]{2003ApJ...595L.107E}, the hard X-ray flux $I_o(\epsilon)$ from ROI-0 in each energy range is calculated by integrating over the FWHM area, while the spectra $I_1(\epsilon)$ and $I_2(\epsilon)$ from the weaker regions ROI-1 and ROI-2 are obtained by integrating the images over the entire areal extent of each respective region. We find that ROI-2 accounts for $\sim$$30$\% of the total emission, while ROI-1 is rather faint, accounting for less than 10\% of the total. Indeed, the HXR flux that appears to originate from ROI-1 in these reconstructed images is likely to be due mostly to a number of artifacts associated with pulse pileup, incomplete removal of side lobes from the intense compact source, and the limited number of spatial Fourier-transform components (``visibilities'') measured with \emph{RHESSI} \citep{2002SoPh..210..101H}. Thus, the already small inferred ROI-1 HXR fluxes $I_1(\epsilon)$ shown in Figure~\ref{fig:xspectra} represent \textit{upper limits} to the actual HXR intensity from that region, so that

\[
\frac{I_{ROI-1}}{I_{\rm total}} \leq 0.1 \,\,\, .
\]
Using the thin-target spectral fit to the spatially-integrated HXR spectrum in Figure~\ref{fig:hxr_fit}, the mean electron flux spectrum $\langle n_p V {\cal F}_o \rangle \, (E_o)$ for the entire flare is $\simeq 2 \times 10^{56}$~electrons~cm$^{-2}$~s$^{-1}$ above reference energy $E_o = 20$~keV. Given the considerations of the previous paragraph, an upper limit for the mean source electron flux above 20~keV from ROI-1 is 10\% of this value, i.e., $\langle n_p V {\cal F}_o \rangle_{ROI-1} \lapprox 2 \times 10^{55}$~electrons~cm$^{-2}$~s$^{-1}$. Using the cube of the source FWHM to estimate the volume of ROI-1 gives $V \simeq 1.6 \times 10^{27}$~cm$^3$, which agrees very well with the value $V=1.7 \times 10^{27}$~cm$^{3}$ obtained by \cite{2022Natur.606..674F} from radio observations. Substituting these values with $\delta = 2.92$ and $v_o = 8.4 \times 10^9$~cm~s$^{-1}$ in Equation~(\ref{eq:nnth-np-V}) gives

\begin{equation}\label{nnth-np}
n_{nth} \, n_p \, \lapprox \, 0.79 \times \frac{2 \times 10^{55}}{(8.4 \times 10^{9}) \, (1.6 \times 10^{27})} \simeq 1.2 \times 10^{18}~{\rm cm}^{-6}\, .
\end{equation}

EUV images from SDO/AIA allow us to make an independent estimate the density in ROI-1. Dividing the total emission measure $n_p^{~2} V$ of the EUV-emitting plasma in this region by the previously-determined source volume, we find that the ambient proton density in this region (right panel of Figure~\ref{fig:hxr-50}) is $n_p \simeq 10^{10}$~cm$^{-3}$, i.e.,  consistent with densities commonly found in the preflare corona.

With this constraint in mind, we now consider the implications of Equation~(\ref{nnth-np}).

\begin{enumerate}

\item \citet{2022Natur.606..674F} claim that all of the electrons in ROI-1 are accelerated, i.e., $n_{nth} \simeq 10^{10}$~cm$^{-3}$, and moreover (their Figure~2c) that the number density of thermal electrons is depleted in ROI-1. However, due to quasi-neutrality of a plasma, the proton density must be $n_p \simeq n_{nth}$; Equation~(\ref{nnth-np}) then shows that the number of non-thermal electrons cannot exceed $10^{9}$~cm$^{-3}$. Extended Data Figure~3 of \citet{2022Natur.606..674F} shows a best-fit thermal plasma density of $10^{8.3}$~cm$^{-3}$ (with an uncertainty of an order of magnitude or so) in their sample ROI-1 pixel. Extended Data Figure~5b from the same \cite{2022Natur.606..674F} work uses SDO/AIA data to conclude that the thermal number density $n_{th}$ in ROI-2 has a value of order $10^{10}$~cm$^{-3}$. This value is consistent with our own estimates of the thermal density in that region (right panel of our Figure~\ref{fig:hxr-50}), and we would also note that our Figure~\ref{fig:hxr-50} shows the SDO/AIA-inferred thermal densities in ROI-1 and ROI-2 to be comparable. The thermal density in ROI-1 inferred from SDO/AIA data (our Figure~\ref{fig:hxr-50}) is some two orders of magnitude greater than the radio-observation-based estimates of the same quantity shown in Figure~2c of \cite{2022Natur.606..674F}.

\item If the average ambient (thermal) plasma density in ROI-1 is $n_p=10^{10}$~cm$^{-3}$, as evidenced by the SDO/AIA observations (right panel of Figure~\ref{fig:hxr-50}), then $n_{nth}/n_p \simeq 0.012$, a value that is consistent with other estimates \citep[e.g.,][]{2013A&A...551A.135S} and kappa-like electron distribution in hard X-ray coronal sources \citep{2009A&A...497L..13K,2013ApJ...764....6O} and magnetic reconnection outflow regions \citep{2019ApJ...872..204B}.
\end{enumerate}

We believe that $n_{nth}/n_p \simeq 0.01$ in region ROI-1 is consistent with all the available observations of this flare and with reasonable expectations for the physical nature of ROI-1. The value of $n_{nth}(>20\text{keV})$ obtained from this reasoning can also be corroborated by appealing to thick-target \citep{1971SoPh...18..489B} modeling of the event. A~thick-target fit to the spatially-integrated hard X-ray spectrum in Figure~\ref{fig:hxr_fit} gives an injected electron rate $\dot{N}\simeq 9 \times 10^{35}$~electrons~s$^{-1}$ above $20$~keV. In a comparison of the electron rate at the loop top and at the footpoints, \cite{2013A&A...551A.135S} showed that the coronal $\dot{N}$ is on average $\sim 1.7-1.8$ times larger than the rate of electron precipitation toward the loop footpoints, with this difference resulting from some form of coronal trapping (e.g., turbulent scattering or magnetic mirroring). Applying this correction factor to the thick target $\dot{N}$ gives\footnote{It should be noted that the event in question was a West limb event, with no obvious footpoint sources visible. It is possible that the coronal density is high enough that the corona acts as a thick target \citep{2004ApJ...603L.117V}, effectively stopping the electrons before they reach the chromosphere.  However, if the footpoint sources were indeed occulted by the solar photosphere, the observed HXR intensity does not represent the entire HXR emission, and thus the precipitation rate $\dot{N}_{LT}$, and hence the inferred value of $n_{nth}$, is higher than that obtained from the observed HXR flux.} a ``loop-top acceleration rate'' $\dot{N}_{LT} = n_{nth} \, v_o \, A \simeq 5 \times 10^{35}$~electrons~s$^{-1}$, corresponding to the rate at which electrons are injected from the loop-top acceleration region into the surrounding target. Assuming that the cross-sectional area of the acceleration region is at least $A=8''\times 8'' \simeq 3.3 \times 10^{17}$~cm$^2$, and using $v_o \simeq 8.4 \times 10^9$~cm~s$^{-1}$  (corresponding to the low energy cut-off electron energy $E_o = 20$~keV), gives $n_{nth} < 2 \times 10^8$~cm$^{-3}$, which is much lower than the ambient plasma density from SDO/AIA and gives $n_{nth}/n_p < 0.02$, consistent with the results from thin-target modeling of the ROI-1 region alone.

\begin{figure*}[pht]
    \centering
     \includegraphics[width=0.32\linewidth]{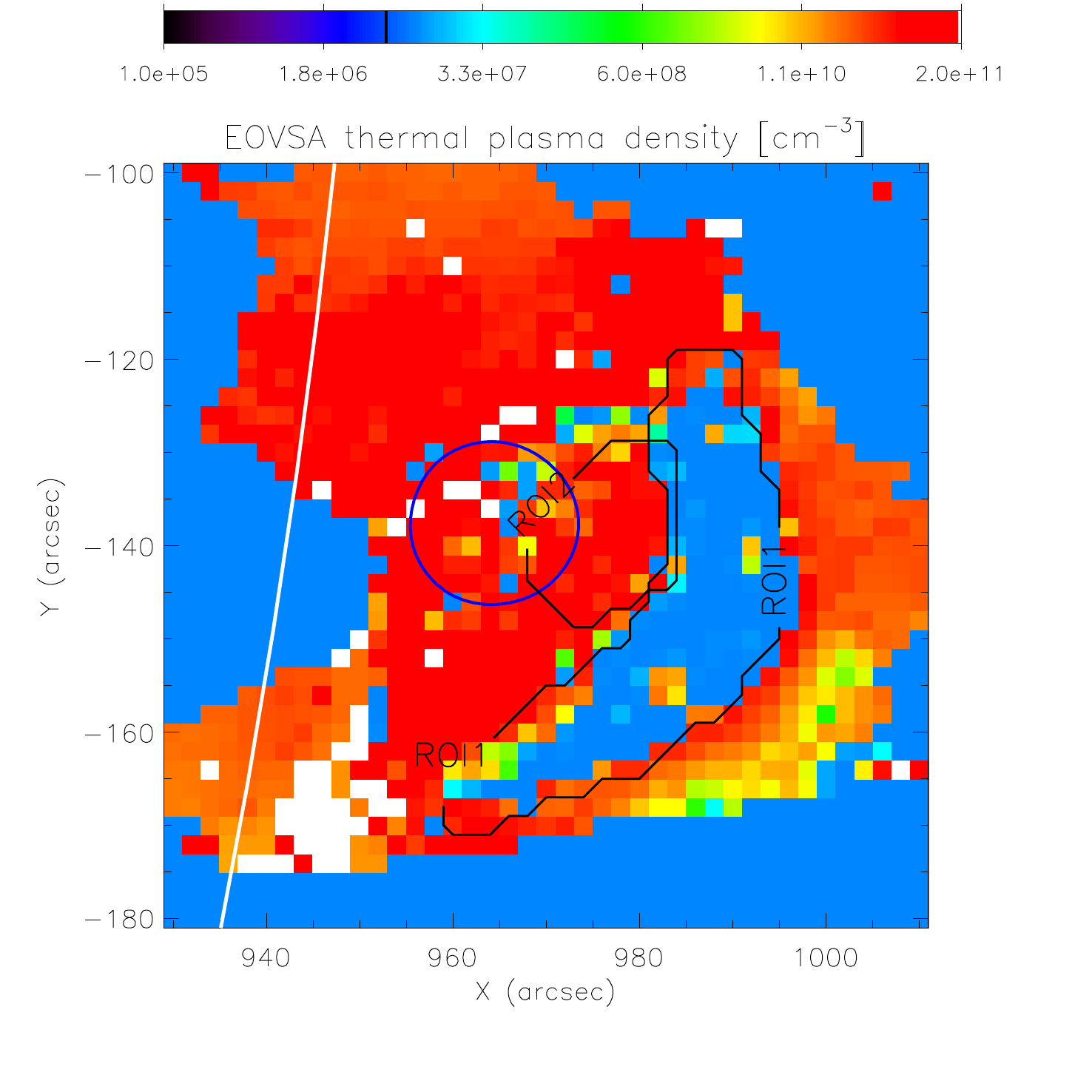}
     \includegraphics[width=0.32\linewidth]{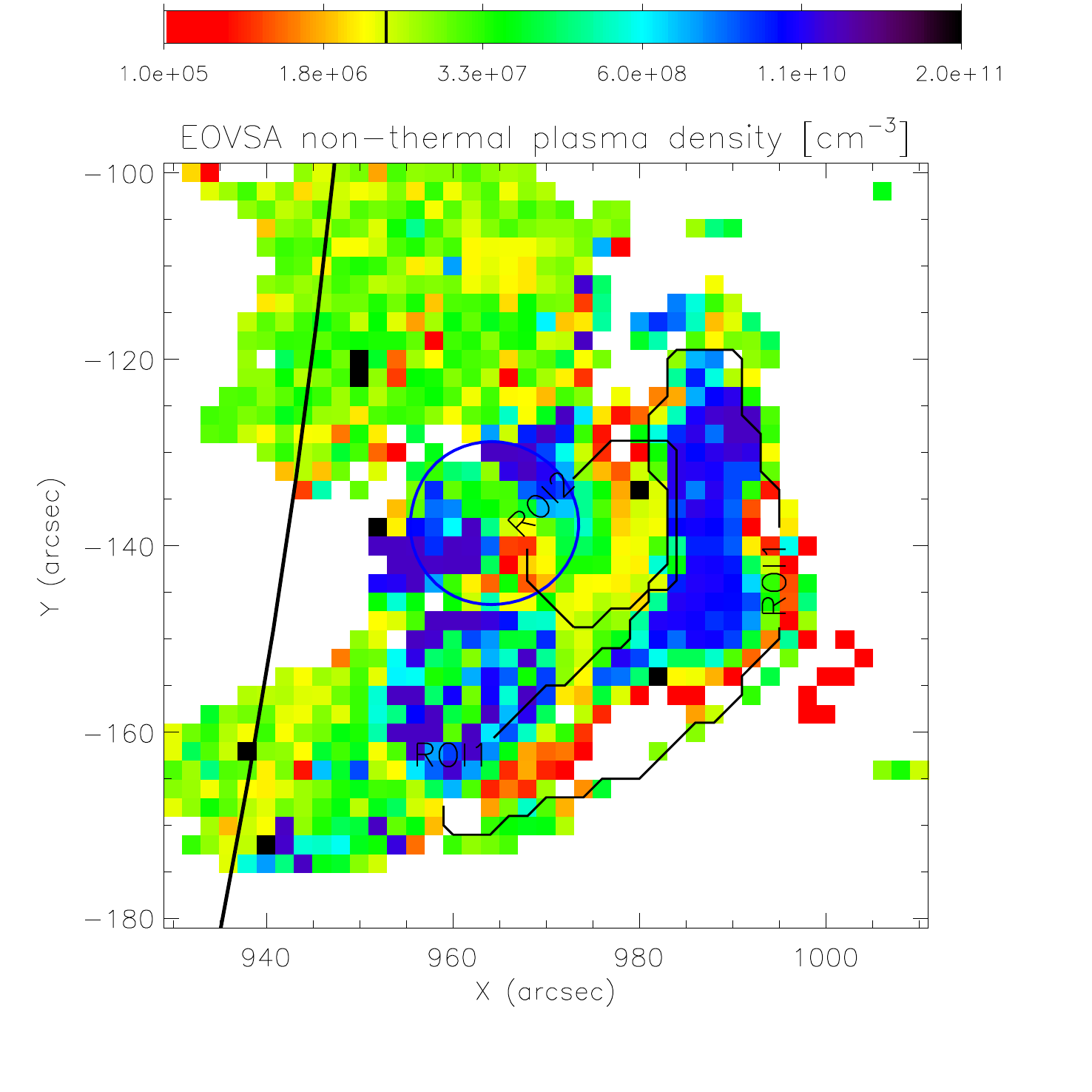}
     \includegraphics[width=0.32\linewidth]{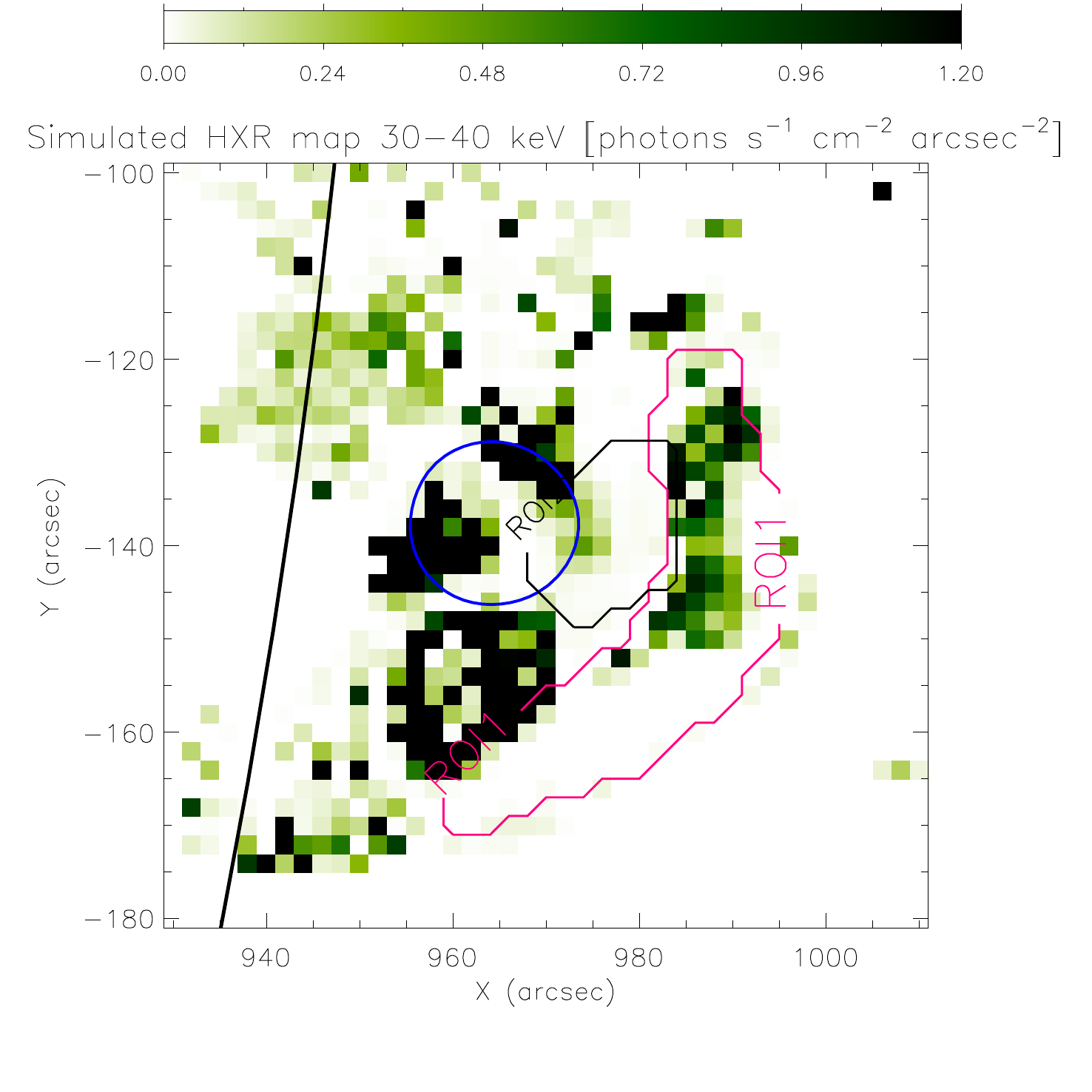}
    \caption{Thermal plasma (left panel) and non-thermal (above 20 keV, center panel) electron number densities as inferred by \citet{2022Natur.606..674F}. The right panel shows the corresponding simulated HXR emission, while the blue contour shows the 50\% level ($0.6$~photons~cm$^{-2}$~sec$^{-1}$~arcsec$^{-2}$) obtained from \emph{RHESSI} imaging spectroscopy observations in the $30-40$~keV range over the same time interval used for Figure~\ref{fig:hxr_fit}.}
    \label{fig:hxr_sim_map}
\end{figure*}

As a final self-consistency check, the nonthermal and thermal electron distribution maps reported by \cite{2022Natur.606..674F} can be used in a thin-target model, \textit{without any additional assumptions}, to produce simulated HXR maps. The simulated hard X-ray emission map in the range $30-40$~keV is shown in the right panel of Figure~\ref{fig:hxr_sim_map}, together with the 50\% contour of the HXR emission observed by \emph{RHESSI} in the same energy range. This panel shows that the spatial distribution of simulated HXR emission based on the on the electron distribution maps of \cite{2022Natur.606..674F} is not consistent with that observed by \emph{RHESSI}. Specifically, the observed HXR flux between $30-40$~keV within the 50\% contour level (Figure~\ref{fig:hxr_sim_map}) is about $0.6$~photons~sec$^{-1}$~arcsec$^{-2}$~cm$^{-2}$, but the nonthermal and thermal electron densities inferred by \cite{2022Natur.606..674F} would predict a significantly higher HXR count rate in many pixels of ROI-1. In fact, the simulated flux for some $2'' \times 2''$~pixels exceeds the observed emission from the entire flare volume.

We further note that the radio spectrum analysis by \cite{2022Natur.606..674F} is performed over $2''$ pixels. This scale is well below the EOVSA beam resolution of $(45-5)''$ for the $(2-18)$~GHz range, so that their non-thermal electron maps must be interpreted with a considerable degree of caution. The required angular resolution ($1''$ at 15 GHz) with better sensitivity, and unprecedented u-v coverage could, however, be achieved by the Square Kilometre Array (SKA) that is under construction \citep{2019AdSpR..63.1404N}, and we encourage such observations\footnote{The Frequency Agile Solar Radiotelescope (FASR), which is currently under development \citep{2022arXiv221010827G}, could also be very useful in this regard.}, ideally in concert with simultaneous HXR imaging spectroscopy measurements (e.g., from the STIX instrument on Solar Orbiter).

\section{Summary and discussion}\label{summary}

The results of the previous section, based on \emph{RHESSI} hard X-ray and SDO/AIA observations of the 2017~September~10 flare, show a discrepancy with the results presented by \cite{2020Sci...367..278F}. Our analysis indicates that the ratio of nonthermal electrons to ambient electrons in ROI-1 at a time near the peak of the X-ray emission is $n_{nth}/\, n_p \simeq 0.01 - 0.02$, while the analysis by \citet{2022Natur.606..674F} using radio observations alone suggests that $n_{nth}/\, n_p \simeq 1$.  In light of these widely discrepant values of $n_{nth}/n_p$ (or, equivalently, the significant excess of HXR emission predicted by the model of \cite{2022Natur.606..674F}, compared with the actual \emph{RHESSI} observations of the same event --- Figure~\ref{fig:hxr_sim_map}), it is worthwhile to revisit the arguments of \cite{2020Sci...367..278F} regarding the values of the physical parameters that characterize the primary reconnection process.

First, we remark that the mean kinetic energy per electron $\langle E\rangle=(1/n_{nth}) \, \int_{E_o}^ \infty E \, n(E) \, dE$ inferred by \cite{2022Natur.606..674F} for ROI-1 corresponds to $\sim$$25$~keV or $\langle E\rangle/k_B \simeq 250$~MK, in a region (Figure~\ref{fig:hxr_sim_map}) where the thermal electron density is claimed to be negligible. This vastly exceeds commonly-accepted values of $20-30$~MK for most flare bulk plasma components \citep[see, e.g.,][for reviews]{1979ApJ...233L.157D,1988SoPh..118...49D,2011SSRv..159..107H,2017LRSP...14....2B,2017ApJ...836...17A}, and even exceeds that of the so-called ``super-hot'' component at a temperature of $30-50$~MK that has been inferred in some events \citep{1985SoPh...99..263L,1996AdSpR..17d..39P,2014ApJ...781...43C}. The bulk energization of a volume of solar plasma to such high equivalent temperatures therefore represents an unprecedented situation.

Second, \cite{2022Natur.606..674F} use Faraday's law $\nabla \times {\cal E} = (1/c) \, \partial B/\partial t$ and write $\nabla \times {\cal E} \simeq {\cal E}/\ell$, where $\ell$ is a characteristic scale length associated with the gradient in the reconnecting magnetic field region. This gives\footnote{the factor 300 converts the cgs statvolt units into Volts} $\cal{E} \simeq$~(300 $\ell/c$) $\partial B/\partial t = (\ell/10^8) \, \partial B/\partial t$~V~cm$^{-1}$.  Using a value $\partial B/\partial t \simeq 5$~G~s$^{-1}$, based on observations reported by \cite{2020Sci...367..278F}, and further estimating that $\ell = 3.65 \times 10^8$~cm (corresponding to $5^{\prime \prime}$~on the solar disk), \cite{2022Natur.606..674F} obtain ${\cal E} \simeq 20$~V~cm$^{-1}$, more than five orders of magnitude greater than the \cite{1959PhRv..115..238D} field ${\cal E}_D \simeq 4 \times 10^{-4} \, (n/10^{10}~{\rm cm}^{-3}) \, (10^6 \, {\rm K} / T) \simeq 1.3 \times 10^{-4}$~V~cm$^{-1}$ (for a density $n = 10^{10}$~cm$^{-3}$ and a temperature $T = 3 \times 10^6$~K). Under the influence of such a large electric field, all of the electrons in the ambient Maxwellian distribution would be accelerated, and they would reach an energy of 20~keV over a very short distance $L \simeq 10^3$~cm, some five orders of magnitude less than the assumed reconnection scale $\ell$. This is not consistent with the claim of \cite{2022Natur.606..674F} that ``this strong super-Dreicer field must be present over a substantial portion of ROI-1.'' Such a value of $L$ also corresponds to an acceleration time $L/v \simeq 10^{-7}$~s and thus, with an accelerated fraction of unity, to a \textit{specific acceleration rate}  \citep[fraction of electrons accelerated to 20~keV per unit time;][]{2008AIPC.1039....3E,2012ApJ...755...32G,2013ApJ...766...28G}
$\eta$~(20~keV) $\simeq 1/10^{-7} = 10^7$~s$^{-1}$, nine orders of magnitude greater than previously inferred for other flares \citep{2013ApJ...766...28G}.

Significant runaway acceleration of the electrons in the ambient Maxwellian will, however, occur for electric field strengths ${\cal E}$ that are merely of order the Dreicer field, much lower than the field strength claimed by \cite{2022Natur.606..674F}.  Furthermore, once this runaway acceleration commences, the fundamental electrodynamic properties in the acceleration region (including the replacement of accelerated particles by a co-spatial return current) will change sufficiently that further buildup of the electric field is not necessary and, indeed, is unlikely. Using an illustrative reconnection scale length value $\ell = 10^3$~cm and the same 5~G~s$^{-1}$ rate of change of magnetic field inferred by \cite{2020Sci...367..278F}, the resulting induced electric field $\cal{E}$ $\simeq 5 \times 10^{-8} \, \ell \simeq 5 \times 10^{-5}$~V~cm$^{-1}$, approximately one-third of the Dreicer field. Such a field will cause  runaway acceleration of electrons with $v \gapprox v_{crit} = \sqrt{{\cal E}_D/{\cal E}} \, v_{th} \simeq \sqrt{3} \, v_{th}$, representing a fraction $0.5 \times {\rm erfc}(\sqrt{3}/\sqrt{2}) \simeq$~5\% of the ambient Maxwellian distribution. However, it is very likely that the acceleration region is highly inhomogeneous, so that the required electron energies are reached through a more stochastic process involving a succession of smaller impulses acting in different directions and with different efficiencies, so that electrons are accelerated through a Fermi acceleration process \citep[e.g.,][]{1997JGR...10214631M,2012ApJ...754..103B,2015ApJ...814..137Z,2020ApJ...902..147G,2021PhRvL.126m5101A}, rather than as the result of a single large-scale unidirectional acceleration event.

With 5\% of the electron population undergoing runaway acceleration to deka-keV energies at any given time, and a collisional repopulation of the tail on a timescale of $0.5$~s, the \textit{instantaneous} number density of accelerated electrons, $n_{nth}$, corresponding to the number density of electrons in the high-energy runaway tail of the distribution, is a relatively small fraction of the ambient number density $n_p$. Further, the associated value of the specific acceleration rate is $\eta$~(20~keV) $\simeq 0.05/0.5 = 0.1$~s$^{-1}$, consistent with the \emph{RHESSI} and SDO/AIA observations of Section~\ref{HXR_obs} and comparable to the specific acceleration rates inferred previously using different methods \citep[e.g.,][]{2013ApJ...766...28G}.  In contrast, a much larger overall population of electrons is successively accelerated over significantly longer timescales. For an ambient density $n_p \simeq 10^{10}$~cm$^{-3}$ and a source volume $V \simeq 10^{27}$~cm$^3$, the total number of available electrons is $\simeq 10^{37}$, so that the electron acceleration rate of $9 \times 10^{35}$~s$^{-1}$ inferred from thick-target modeling of the event (Section~\ref{HXR_obs}) corresponds to an acceleration of all the electrons in the corona in $\sim 10$~s, shorter than the duration of the HXR burst. This constitutes the well-known ``number problem'' \citep[e.g.,][]{1971SoPh...18..489B} that requires a continual replenishment of the acceleration region, e.g., by a cospatial return current carried by ambient thermal electrons \citep{1977ApJ...218..306K,1980ApJ...235.1055E,Alaoui_2017}.

The dramatically different values for $(n_{nth}/ \, n_p)$, from $\simeq$$0.01$ to $\simeq$$1$, clearly presents us with a dilemma. How are such disparate results to be reconciled? One possibility lies in the fact that the emitted microwave flux is rather insensitive to the value of the low-energy cutoff energy $E_o$ \citep[see Figure~1 in][]{2003ApJ...586..606H}. Further, Extended Data Figure~3 of \cite{2022Natur.606..674F} claims a relatively high spectral index $\delta \simeq 5-6$ for a ``typical'' pixel in ROI-1, much higher than the value $\delta \simeq 3$ obtained from the total spatially integrated HXR spectrum (Figure~\ref{fig:hxr_fit}). This indicates \citep[probably similarly to][]{2021ApJ...908L..55C} that such a high value of $\delta$ is not applicable to the deka-keV regime from which most of the contribution to $n_{nth}$ arises. Such a flattening of the electron spectrum at lower energies would reduce the value of $n_{nth}$ from that claimed by \cite{2022Natur.606..674F}. Indeed, since the total number of accelerated electrons ($\int_{E_o}^\infty E^{-\delta-1/2} \, dE \propto E_o^{1/2-\delta}$; Equation~(\ref{eq:nE})) depends rather sensitively on the the values of both $E_o$ and $\delta$, it is possible that a much lower value of $n_{nth}$, comparable with that deduced from HXR and EUV observations, is also consistent with the observed microwave emission. Interestingly, the self-consistent simulations of electron acceleration during magnetic reconnection in a macroscale system \citep{2021PhRvL.126m5101A}, as well as spatially-extended turbulent electron acceleration \citep{2023arXiv230113682S}, also suggests that the instantaneous number density of nonthermal electrons remains small.

\begin{acknowledgements}

The authors are thankful to G. Fleishman for sharing data and useful discussions. We also thank the referee for a thorough review of the manuscript and for some excellent suggestions for improving it. This work was supported by STFC consolidated grant ST/T000422/1. AGE was supported by NASA Kentucky under NASA award number 80NSSC21M0362. GGM was supported by grant 21-16508J of the Grant Agency of the Czech Republic, the project RVO:67985815, the project LM2018106 of the Ministry of Education, Youth and Sports of the Czech Republic, and the State Assignment 0040-2019-0025. \emph{RHESSI} data are available freely from \url{https://hesperia.gsfc.nasa.gov/rhessi3/}

\end{acknowledgements}

\bibliographystyle{aasjournal}
\bibliography{all_electrons}

\end{document}